\documentclass[12pt]{article}

\usepackage{graphicx}
\usepackage{subcaption}
\usepackage{siunitx}
\usepackage{authblk}

\usepackage{geometry}
\usepackage{setspace}
\title{\Large\textbf{ Structural and chemical mechanisms governing stability of inorganic Janus nanotubes}}

\author[1]{\small Felix Tim B\"olle}
\author[1]{\small August E. G. Mikkelsen}
\author[2]{\small Kristian S. Thygesen}
\author[1]{\small Tejs Vegge}
\author[1,*]{\small Ivano E. Castelli}

\affil[1]{%
Department of Energy Conversion and Storage, Technical University of Denmark, Anker Engelundsvej 411, DK-2800 Kgs. Lyngby, Denmark.}

\affil[2]{%
Department of Physics, Technical University of Denmark, Fysikvej 309, DK-2800 Kgs. Lyngby, Denmark.}

\affil[ *]{\textit{Email: ivca@dtu.dk}}

\begin{document}
\maketitle
\newpage

\begin{centering}
\section*{Abstract}
\end{centering}
One-dimensional inorganic nanotubes hold promise for technological applications due to their distinct physical/chemical properties, but so far advancements have been hampered by difficulties in producing single-wall nanotubes with a well-defined radius. In this work we investigate, based on Density Functional Theory (DFT), the formation mechanism of 135 different inorganic nanotubes formed by the intrinsic self-rolling driving force found in asymmetric 2D Janus sheets.
We show that for isovalent Janus sheets, the lattice mismatch between inner and outer atomic layers is the driving force behind the nanotube formation, while in the non-isovalent case it is governed by the difference in chemical bond strength of the inner and outer layer leading to steric effects. From our pool of candidate structures we have identified more than 100 tubes with a preferred radius below 35 \si{\angstrom}, which we hypothesize can display unique properties compared to their parent 2D monolayers. Simple descriptors have been identified to accelerate the discovery of small-radius tubes and a Bayesian regression approach has been implemented to assess the uncertainty in our predictions on the radius.


\section*{Introduction}
In the last decades miniaturization of devices has been a main trend driving the electronics industry. In addition to reducing the usage of raw materials, nanomaterials often show improved properties compared to their larger counterparts. Among these nanomaterials are two-dimensional (2D) sheets, one-dimensional (1D) structures such as nanotubes and nanoribbons, and zero-dimensional (0D) nanoparticles.

Since their discovery, nanotubes have shown promise for a wide range of applications including gas separation and capture, catalysis, solid lubrication and controlled drug delivery.\cite{serra2019overview} In addition to the well-known carbon nanotubes \cite{iijima1991helical} numerous inorganic nanotubes have been synthesized .\cite{Tenne1992, rao2003inorganic} Although the first successful synthesis of single-wall MoS$_2$ nanotubes has been reported [5], such structures usually appear together with numerous multi-wall tubes showing a distribution of radii and wall thicknesses.\cite{seifert2002stability} These multi-wall structures alleviate the built-in strain energy through van der Waals interactions in between the layers leading to an increase in stability.\cite{serra2019overview} Overall, this has made it difficult to establish an experimental synthesis pathway to produce single-wall tubes with a specific radius and controllable physio-chemical properties.

A possible solution to this problem is the approach of considering asymmetric sheets, which can naturally wrap and form nanotubes with a well defined size. Due to the asymmetry, the unsupported sheet is expected to be unstable compared to other curled shapes, such as tubes or scrolls. Pauling, already in the 1930s, mentioned that the driving force of sheets to curve is related to the lattice-mismatch between the two inner and outer atomic layers.\cite{pauling1930structure} Single-wall inorganic nanotubes with well-defined diameters hold promise for technological applications, not only because of their reduced dimensionality, but also for their unique properties, often inherently different from the ones of the corresponding asymmetric sheets. An example of a small-radius, single-wall nanotube formed from an asymmetric sheet is imogolite (Al$_2$SiO$_3$(OH)$_4$) which was first discovered in volcanic ash soil \cite{yoshinaga1962imogolite} and later synthesized. \cite{cradwick1972imogolite,farmer1977synthesis} Other tubular minerals include chrysotile (Mg$_3$Si$_2$O$_5$(OH)$_4$) and halloysite (Al$_2$Si$_2$O$_5$(OH)$_4$) that however occur as multi-wall tubes. \cite{bates1950morphology,whittaker1956structure} Besides naturally occurring nanotubes, "misfit-layer" compounds, composed of two separate sheets, make use of the lattice mismatch between the two sheets to induce a natural driving force to form a tube. \cite{panchakarla2014nanotubes}

One of the possible classes of materials forming asymmetric 2D monolayers are Janus sheets, like MoSSe \cite{lu2017janus} or BiTeI \cite{fulop2018exfoliation}, which can be wrapped to form 1D tubes.\cite{zhao2015ultra,oshima2020geometrical} A recent work\cite{zhao2015ultra} has shown that radii well below 35 \si{\angstrom} are needed to create single-wall Janus transition metal dichalcogenide (TMD) tubes, which have significantly different (electronic) properties from the corresponding asymmetric sheet. Although facile synthesis routes for the production of single-wall inorganic nanotubes has long been actively researched, not much attention has been paid to the question of which materials would be able to make such a structure avoiding the creation of multi-wall tubes. Consequently, a high throughput study on the stability of a wide range of Janus-based nanotubes, would provide valuable information for guiding future synthesis of small-radius single-wall nanotubes.

In this work we present a comprehensive screening study in the framework of Density Functional
Theory (DFT) on the stability of 135 different inorganic nanotubes generated from the rolling of asymmetric 2D Janus sheets. The calculations focus on the stability and strain energy of the chosen nanotubes. The total number of DFT relaxations performed in this work amounts to approximately 4500. We show that for pure chalcogen or halogen tubes (isovalent anions), the wrapping mechanism is mostly governed by the lattice-mismatch between the two inner and outer atomic layers, while for mixing anions (non-isovalent anions) this is dominated by the difference in valency between the X/Y elements. These findings provide a physical foundation for designing Janus nanotubes with optimal (small) radii.


\section*{Results}

\begin{figure}[h]
\centering
  \includegraphics[width=1.\linewidth]{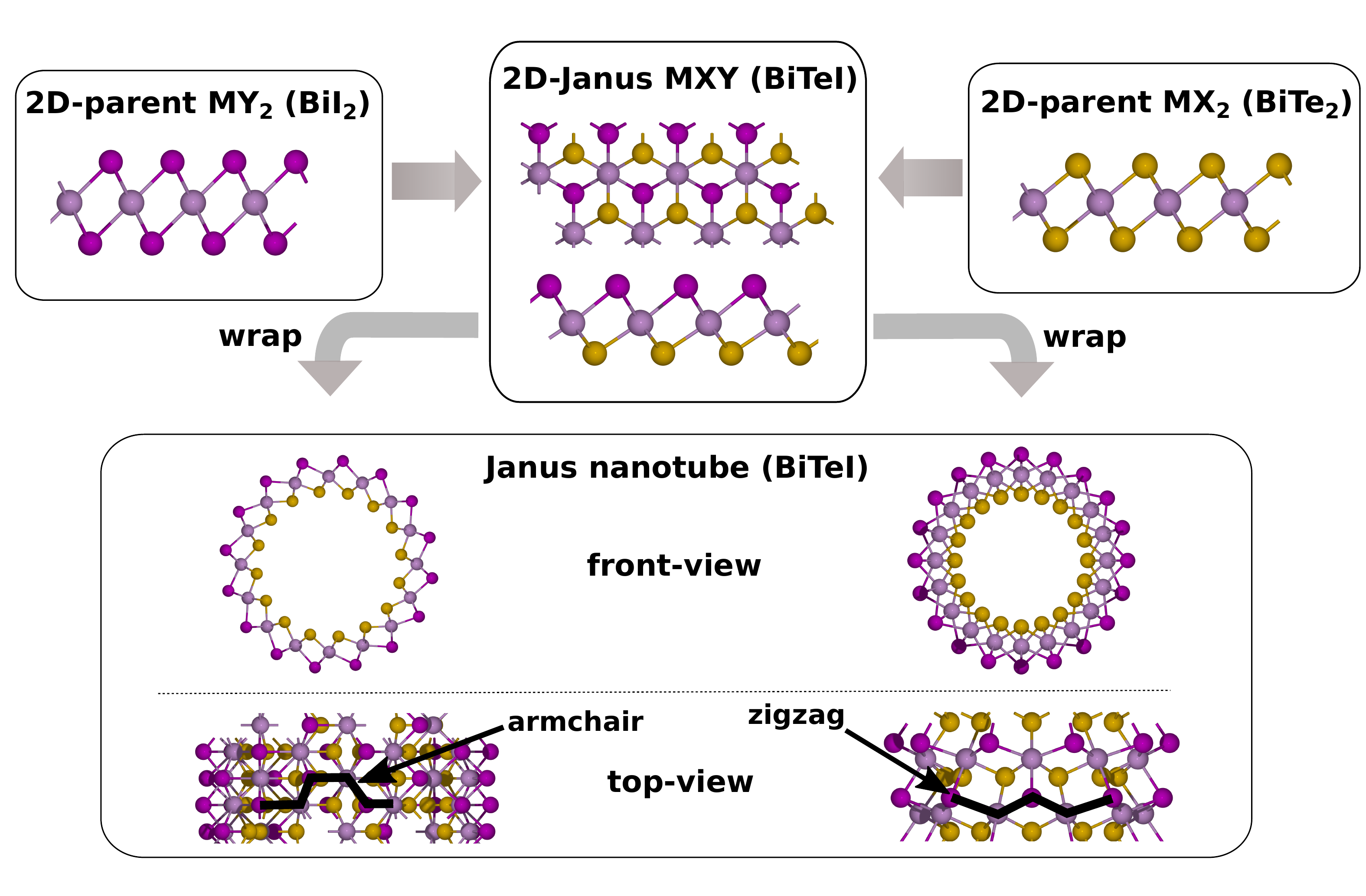}
\caption{The symmetric MY$_{2}$ and MX$_2$ are the parent sheets for the asymmetric 2D Janus MXY sheet, which can be wrapped up to form a 1D Janus nanotube. In this case M = Bi (light purple), X = Te (dark yellow) and Y = I (dark purple).}
\label{fig:wrap_tube_figure}
\end{figure}

The nanotubes considered in this study consist of three layers (MXY, as illustrated in Figure \ref{fig:wrap_tube_figure}) composed of different mid-layer elements (M = \{Ti, Zr, Hf, V, Nb, Ta, Cr, Mo, W, Fe, Ge, Sn, As, Sb, Bi\}) decorated with inner X and outer Y chalcogen and halogen atoms (X,Y = \{O, S, Se, Te, Cl, Br, I\}). Here we denote all group 16 elements including oxygen as chalcogens. For the three pnictogens (As, Sb and Bi) having 3+ as one of their possible oxidation states, we mix chalcogens and halogens in the structures. For the remaining 12 elements, the inner X and outer Y elements are either chalcogens or halogens. The idea of mixing chalcogens and halogens to form 2D MXY Janus sheets was recently explored in Ref. \cite{riis2019classifying} , but has so far not been pursued in the context of nanotubes. We construct the tubes by rolling up 2D layers in both the T- and H-phase crystal structures, corresponding to the crystal structures found for the experimentally synthesized MoSSe \cite{lu2017janus} and BiTeI \cite{fulop2018exfoliation} 2D sheets, along both the armchair and zigzag directions.

\subsection*{Nanotube strain energy}
Two main quantities that are needed to characterize an asymmetric nanotube are the optimal radius, which defines the most stable nanotube size, and the strain energy, which defines the energy associated with the wrapping of a 2D sheet into a nanotube (negative strain energies indicate a spontaneous wrapping).

The strain energy is defined as the difference between the energy of the nanotube and the corresponding 2D sheet.\cite{tersoff1992energies} In formula:
\begin{equation}
    E_{strain}(R) = \frac{E_{tube}}{N_{tube}} - \frac{E_{MXY}}{N_{MXY}}
\end{equation}

, where $E_{tube}$ is the energy of a nanotube with $N_{tube}$ atoms and $E_{MXY}$ is the energy of the corresponding 2D Janus sheet with $N_{MXY}$ atoms in the unit cell.
In the infinite limit $R \rightarrow \infty$, the strain energy is zero, since the energy per atom of a tube is equal to the energy per atom of an infinite 2D Janus sheet.

It has been shown that for symmetric tubes (carbon, for example) the nanotube strain energy follows a 1/R$^2$ dependence. \cite{tibbetts1984carbon, robertson1992energetics} This relationship does not hold for asymmetric tubes in which the strain energy curve exhibits a minimum.  \cite{guimaraes2007imogolite, d2009single, guimaraes2010structural,zhang2003formation} Instead, it can be more accurately described using the equation: \cite{guimaraes2007imogolite}
\begin{equation}
\label{eq:second_order_model}
    E_{strain}(R) = \frac{a}{R^2} + \frac{b}{R}~.
\end{equation}

\begin{figure}[h]
\centering
  \includegraphics[width=.9\linewidth]{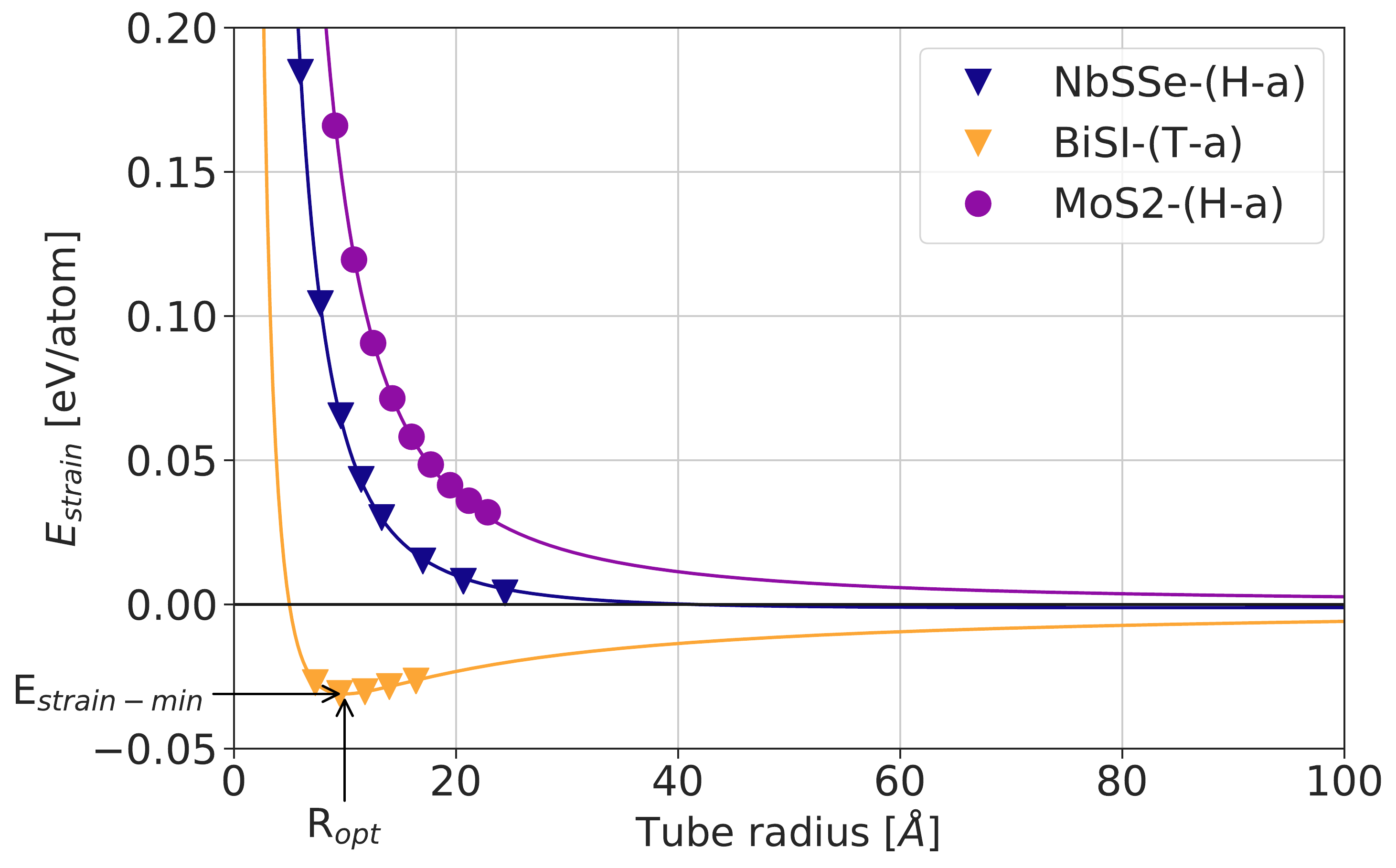}
\caption{Strain energy as a function of the radius for the symmetric MoS$_2$ nanotube and the two asymmetric NbSSe and BiSI nanotubes. The label in brackets indicates the prototype (H-/T-phase) and the wrapping direction (a/z indicating armchair/zigzag).}
\label{fig:inorganic_strains}
\end{figure}

Extrapolating the function using the obtained DFT data and evaluating the function at the minimum strain energy $E_{strain-min}$ leads to the optimal tube radius $R_{opt}$ (see also Figure \ref{fig:inorganic_strains}).
We note that, although Eq. \ref{eq:second_order_model} fits well in the region around the optimal tube radius, large strain energies can lead to a deviation. \cite{demichelis2016serpentine} We take this into consideration during the screening using Bayesian statistics (details in the SI) which helps to identify cases where the function chosen does not capture the observed data points well across all tube radii.

As an example, Figure \ref{fig:inorganic_strains} shows the strain energy as a function of the tube radius for three different materials, comparing a symmetric MoS$_2$ tube with the two studied asymmetric tubes NbSSe and BiSI.
The symmetric MoS$_2$ tube shows a 1/R$^2$ dependence of the strain energy over the tube radius indicating the single-wall nanotube is less stable than the infinite sheet. This is not the case for the asymmetric NbSSe and BiSI tubes, where the strain curve exhibits a minimum. The strain energy curve for BiSI shows a strain energy minimum of -31 meV/atom at the optimal radius of $\sim$ 10 \si{\angstrom} while the strain energy curve for NbSSe is instead very shallow (minimum at -1.1 meV/atom at 85 \si{\angstrom}) due to the minor lattice-mismatch of 0.96 between the two parent sheets NbS$_2$ and NbSe$_2$. Such a shallow strain energy curve makes it difficult to establish an optimal radius.

\subsection*{Stability and optimal radius}

\begin{figure}[h]
\centering
  \includegraphics[width=.625\linewidth]{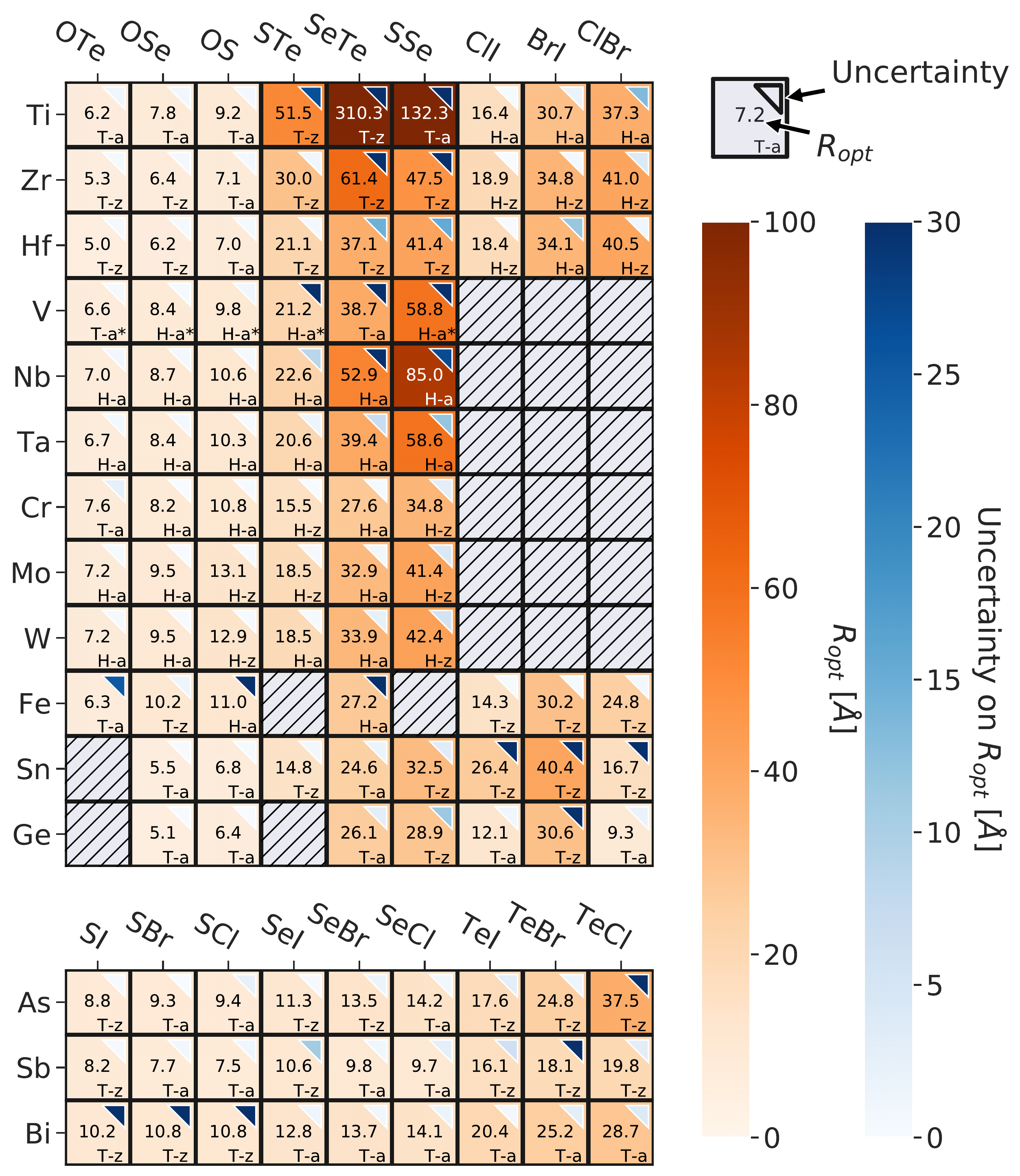}
\caption{Overview of the 135 investigated materials and their extrapolated optimal radius according to equation \ref{eq:second_order_model}. It is ensured that at least three data points after filtering according to the criteria described in the SI exist (otherwise marked with a hatched box). The asterisk indicates that the difference in energy between the H-/T- phase is less than 10 meV/atom and that the smaller optimal tube radius is chosen given the extrapolated optimal radius from both prototypes. The lowercase letters a/z indicate which wrapping direction (armchair/zigzag) is preferred.}
\label{fig:heatmap_radii}
\end{figure}

Figure \ref{fig:heatmap_radii} shows the optimal tube radii for all studied materials and its associated uncertainty. For around 20 structures the uncertainty on the radius is estimated to be larger than 30 \si{\angstrom} (blue shaded triangles in upper right corner in Figure \ref{fig:heatmap_radii}). We employ Bayesian regression (details in the SI) to automatically asses the uncertainty associated with fitting the obtained DFT data to equation \ref{eq:second_order_model}. This makes it possible to spot data points during the screening study that might require a more detailed investigation, while the data points that fit the underlying fitting function show reduced uncertainties. We have identified three situations where the uncertainties are large: (1) The model is not able to fully capture the variation of strain energies across different tube radii. Only few structures, such as SbTeBr, do not fit the proposed model (similar to imogolite, Figure 2 in the SI), leading to large uncertainties in the prediction. (2) The strain energy curves are rather flat due to a less pronounced energy minimum making it difficult to estimate a precise optimal tube radius. This is, for example, the case for NbSSe, TiSSe or VSSe. (3) Too few data points are available, as it happens for BiSI, BiSBr and BiSCl. By looking at the strain energy curve, we note that in these cases adding additional data points to the curve is not necessary, as the optimal radius has already been found using only three data points. However, this leads to poor statistics for the Bayesian regression. In general the values for the optimal tube radii are similar to what has been found in literature for a much smaller set of TMD Janus nanotubes. \cite{oshima2020geometrical}  \\

\begin{figure}[h]
\centering
  \includegraphics[width=.6\linewidth]{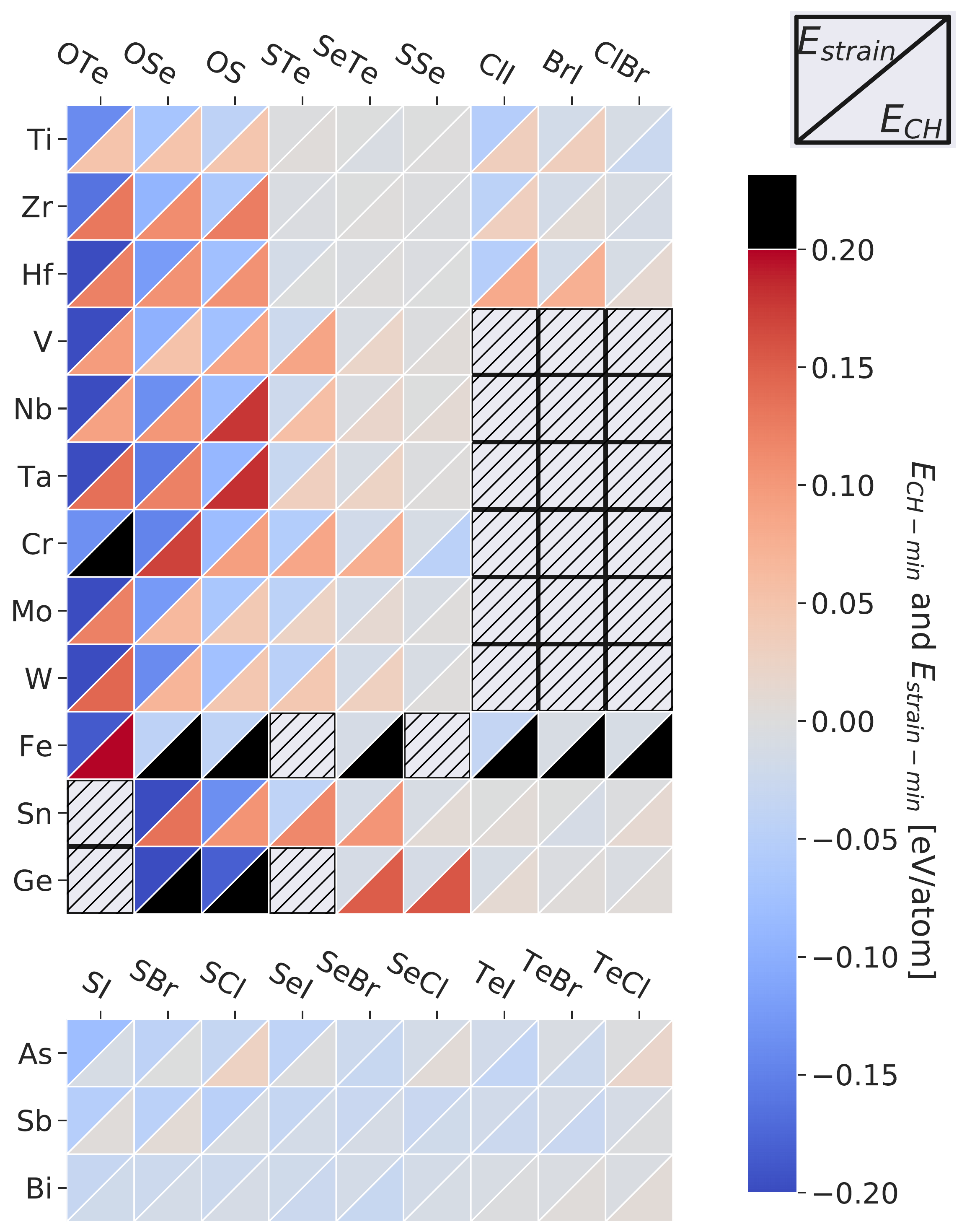}
\caption{The heatmap shows convex hull and tube strain energies at the optimal tube radius. Blue and bright colors indicate good stability, while a red color indicates poor stability. Values exceeding the metastability criterion of 0.2 eV/atom are shown in black.}
\label{fig:heatmap_ch}
\end{figure}

Figure \ref{fig:heatmap_ch} shows the minimum strain energies corresponding to a tube at the optimal tube radius as shown in Figure \ref{fig:heatmap_radii}. To compare the energy of the tube to its most stable 3D bulk structures, we also calculate the convex hull energy of the tube $E_{CH}$ at its optimal radius ($E_{CH-min}$). Except for Fe, all studied nanotubes show good stability against the decomposition into competing bulk structures (taken from the Materials Project database \cite{ong2013python}) for at least one of the calculated combinations (here we define a combination stable when the energy of the candidate compound is within $ 0.2 $ eV/atom above the convex hull to account for a possible metastability \cite{castelli2012computational, sun2016thermodynamic, esposito2020metastability}). This is in good agreement with published studies on the stability of 2D Janus monolayers. \cite{riis2019classifying} Given the mid-layer element is in its preferred oxidation state, the resulting tube shows higher stability compared to the case of an unfavored oxidation state as expected. Ge, for example, generates more stable tubes when combined with two halogens (Ge$^{2+}$) than with two chalcogens (which would need a Ge$^{4+}$, instead). We do not find stable nanotubes for 23 combinations (hatched boxes in Figure \ref{fig:heatmap_ch}). 18 of these materials can be attributed to the transition metal in its unfavored 2+ oxidation state, when paired with two halogens. The remaining 5 materials contain either Fe, Sn or Ge which form in general less stable nanotubes for most of the studied combinations (\textit{i.e.} more than 0.1 eV/atom above the convex hull). Because of their stability in the 3D form, oxygen containing tubes are in general more prone to decompose compared to the pure chalcogenide ones. 

For almost 90 \% of the materials, the energy difference between the armchair and zigzag wrapping direction is below 10 meV/atom, which indicates that there is only a weak driving force causing wrapping up around a specific direction. Although the armchair and zigzag wrapping directions only have a minor impact on the stability of the tube, it can be expected that the wrapping direction has a larger impact on the electronic properties due to the difference in bond distances in these tube configurations. We have recently demonstrated this for MoSTe, but a more comprehensive study focusing specifically on the electronic properties would be needed to establish general design rules. \cite{august2020}

The combination of the three metals As, Sb and Bi mixed together with a chalcogen sitting inside and a halogen element sitting outside of the structure generates stable and small-radius nanotubes with a rather small strain energy associated with its optimal radius. For instance, AsSI has a minimum strain energy of -80 meV/atom which is $\sim$ 40 meV smaller than that of the experimentally observed imogolite nanotube when compared to computational reference data found in literature. \cite{lee2011origin} The MXY nanotubes share with imogolite, which is known to exist as a single-wall nanotube, the shape of the strain energy curve (see BiSI in SI Figure 2).

\begin{figure}[h]
\centering
  \includegraphics[width=.7\linewidth]{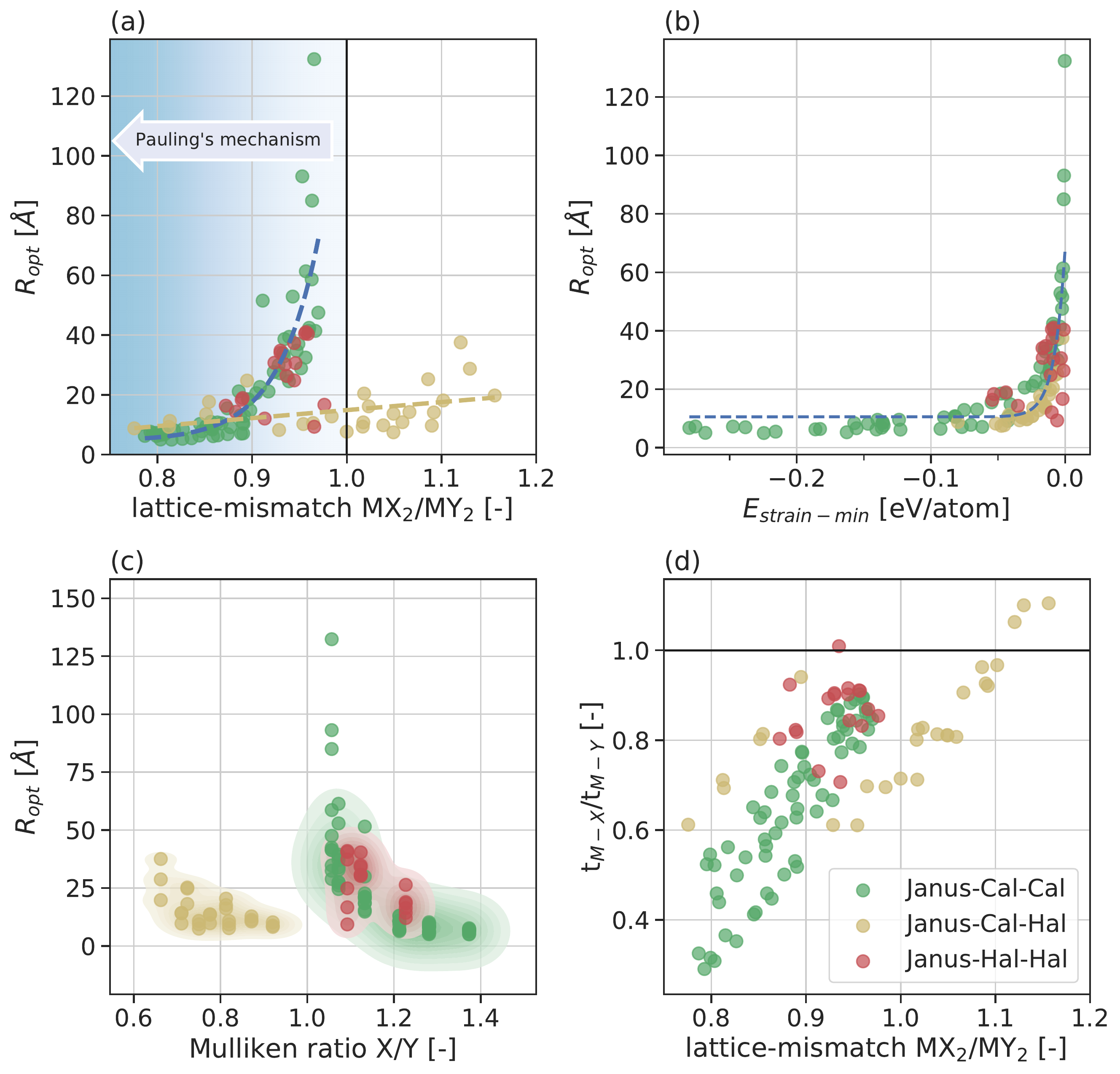}
\caption{R$_{opt}$ versus the lattice mismatch of the corresponding parent sheets (lattice mismatch
 calculated as a$_{MX_2}$/a$_{MY_2}$ with a being the lattice constant) (a), the minimum strain energy at the optimal radius (b) and the Mulliken ratio of the corresponding Mulliken electronegativities of outer and inner elements (Mulliken ratio X/Y) (c) are shown. Figure (d) shows the ratio in layer thicknesses in the 2D Janus sheet versus the lattice-mismatch. The different colors represent the three classes of pairing the mid-layer with either two chalcogens (Janus-Cal-Cal, \{O, S, Se, Te\}, green) , two halogens (Janus-Hal-Hal, \{Br, I, Cl\}, red) or a mix of chalcogens inside and halogens outside (Janus-Cal-Hal, yellow). }
\label{fig:properties_correlated}
\end{figure}

Figure \ref{fig:properties_correlated} reports the correlation between various quantities, namely the extrapolated optimal radius $R_{opt}$ versus the lattice-mismatch a$_{MX_2}$/a$_{MY_2}$ (a), which is the important parameter in Pauling's mechanism, the minimum strain energy $E_{strain-min}$ (b) and the ratio of the Mulliken electronegativities of the X/Y elements (c). Figure \ref{fig:properties_correlated} (d) shows the ratio in layer thickness $t_{M-X}$/$t_{M-Y}$ versus the lattice-mismatch a$_{MX_2}$/a$_{MY_2}$, where the thickness is measured as the M-X and M-Y element distance along the vacuum direction in the 2D Janus sheet (inter-layer distance). A similar plot showing the optimal radius versus the convex hull energies can be found in the SI in Figure 3. Figure \ref{fig:properties_correlated} (b) indicates that very low strain energies are not necessary to obtain small-radius nanotubes. Nanotubes with a diameter smaller than 15 \si{\angstrom} are predicted to be found in a range of strain energy minima from -15 meV/atom (BiSeBr) to -280 meV/atom (TaOTe, which has the most negative strain energy in our dataset). The optimal radius increases sharply when approaching the limit of no lattice-mismatch between the MX$_2$ and MY$_2$ parent sheets (\textit{i.e.} lattice-mismatch a$_{MX_2}$/a$_{MY_2}$ = 1, see Figure \ref{fig:properties_correlated} (a) - blue dashed curve). Additionally, the lattice-mismatch is correlated with the ratio of ionic radii when mixing two chalcogens (Janus-Cal-Cal class following trend OTe $>$ OSe $>$ OS $>$ STe $>$ SeTe $>$ SSe) or two halogens (Janus-Hal-Hal class following ClI $>$ BrI $>$ ClBr) with each other.

The lattice-mismatch between the two parent sheets can give an estimate on the optimal tube radius in the case of isovalent anions, while it fails for mixing halogens with chalcogens (Janus Cal-Hal class). Here, it would would predict the halogen to sit inside of the tube, instead of on the outside, as several of the materials show a lattice mismatch a$_{MX_2}$/a$_{MY_2}$ larger than 1 (Figure \ref{fig:properties_correlated} (a), yellow dashed line). For instance, the MoOTe parent sheets (H-phase) have a lattice mismatch of 0.79 ($a_{MoO_2}$ = 2.82 \si{\angstrom}, $a_{MoTe_2}$ = 3.55 \si{\angstrom}) leading to the small radius of 7.2 \si{\angstrom}. Conversely, the BiTeCl parent sheets (T-phase) have a lattice mismatch of 1.06 since the parent structure BiCl$_{2}$ (a = 3.68 \si{\angstrom}) has a smaller lattice constant than BiTe$_{2}$ (a = 3.9 \si{\angstrom}). Based on the lattice mismatch, the tube should wrap in a way that the Chlorine atoms are inside of the tube. 
The three different classes appear well-separated when the radius is plotted versus the ratio of the Mulliken electronegativities of the X and Y element (Figure \ref{fig:properties_correlated} (c)). Having more electronegative elements sitting on the inside of the tube does not seem to be a necessary criterion for forming small-radius nanotubes (Janus-Cal-Hal, yellow).

\begin{figure}[h]
\centering
  \includegraphics[width=1.\linewidth]{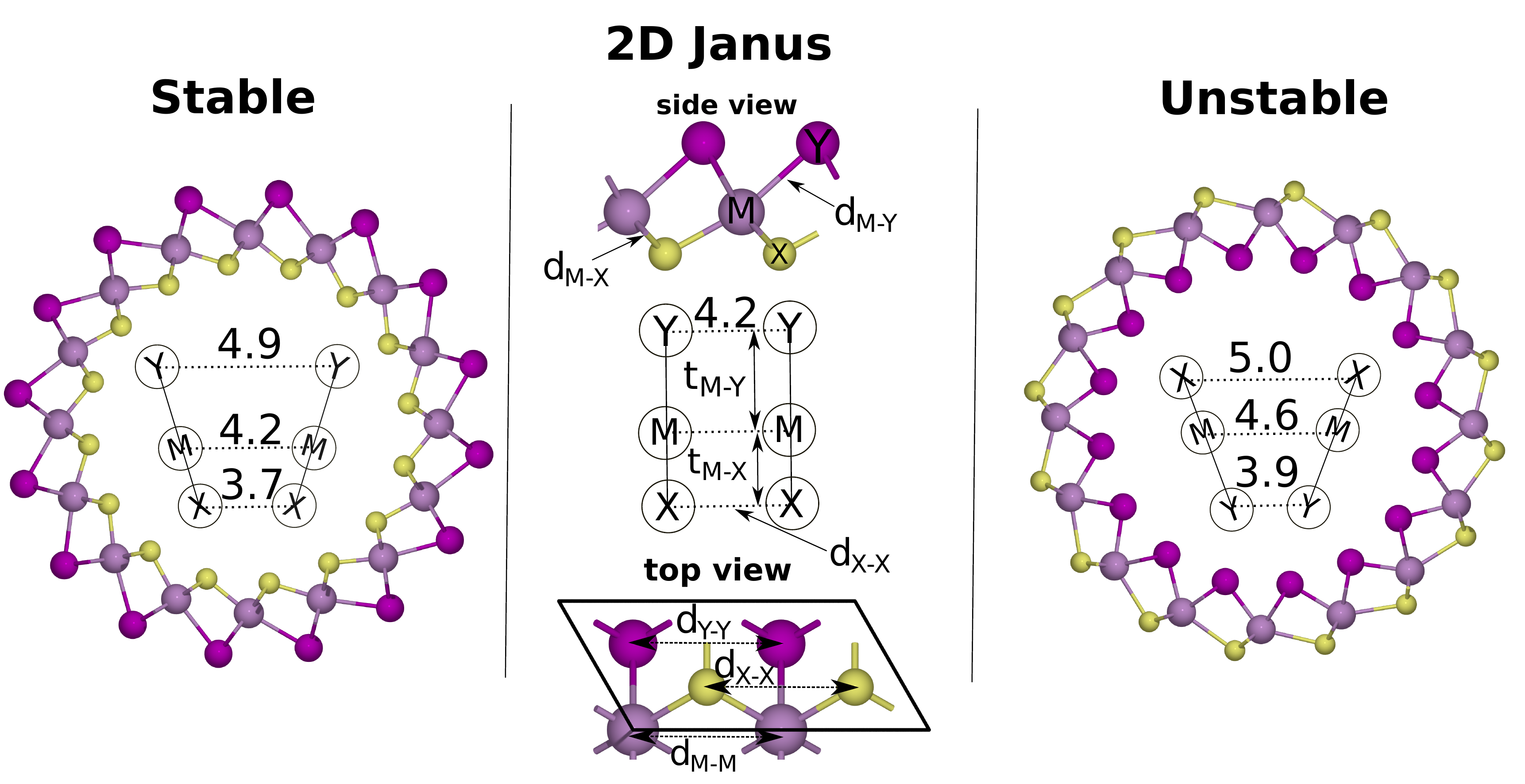}
\caption{Side (or cross-section) view for the energetically favored armchair BiSI nanotube, the infinite 2D Janus layer and the unstable inverse wrapped BiIS nanotube. Measures are given in \si{\angstrom}. }
\label{fig:bond_distances_variation}
\end{figure}

A possible explanation on the reason why the chalcogen is sitting inside of the non-isovalent tube can be made by investigating the bond lengths in the non-isovalent structures closer. Shevelkov \textit{et al.} \cite{shevelkov1995crystal} studied the experimental 3D-bulk crystal structure of layered BiTeI and find the Bi-X bond distance in these structures to be significantly longer (ionic) than the ones found in bismuth trihalides. Additionally, the geometry of the (BiTe) layer is shown to be comparable to the one found in Bi-bulk metal and the bond-distance of Bi-Te in the BiTeI layer is similar to the one found in bismuth tellurides. \cite{shevelkov1995crystal}

To illustrate this, we take the BiSI structure in Figure \ref{fig:bond_distances_variation} and assume the Bi-I (d$_{M-Y}$), Bi-Bi (d$_{M-M}$) and Bi-S (d$_{M-X}$) bond distances to not vary significantly when the 2D Janus sheet is being wrapped up into a tube. By looking at the sheet from the side, the layer thicknesses of both inner and outer layer are $t_{M-X}$ and $t_{M-Y}$, respectively (Figure \ref{fig:bond_distances_variation} mid column). Larger bond-distances now also lead to thicker layers. This means that the M-Y layer is considerably thicker, owed to the ionic like bond between the M-layer element and the halogen. Larger thicknesses of the M-Y layer impose a constraint resulting in M-X bonds taking less space inside the tube and therefore leading to less steric effects as opposed to having M-Y bonds on the inside of the tube (Figure \ref{fig:properties_correlated} (d)). Additionally, we observe for the non-isovalent structures a shortening in M-X bond lengths upon wrapping the infinite 2D Janus sheet into a nanotube leading to the effective thickness of the MX layer in a tube being even thinner than in the sheet.

In order to further clarify this point we compare the stable BiSI tube and the energetically unstable inverse wrapped counterpart BiIS in Figure \ref{fig:bond_distances_variation} on the right. In the inverse wrapped case steric effects in between I-I elements even lead to a bond expansion of the Bi-Bi bond, d$_{M-M}$, from 4.2 \si{\angstrom} to 4.6 \si{\angstrom}. Given both tube configurations differ by 0.14 eV/atom in energy in favor of the tube in which the chalcogen element sits inside, the importance of the (BiTe) layer for the stability is underlined. Plotting the stability difference in between the stable (chalcogen inside) and the inverse wrapped tube (chalcogen outisde) for nine selected non-isovalent tubes we find the points to follow a positive linear correlation with the ratio in thickness r = $t_{M-X}/t_{M-Y}$ (Figure 4 in the SI).
We conclude that the steric effects caused by a difference in layer thickness significantly influences the stability of the tubes. Steric effects can also explain the reduced minimum strain energy for BiXI $<$ BiXBr $<$ BiXCl structures, as well as the reduced optimal radius for BiSY $<$ BiSeY $<$ BiTeY structures. In these cases, the element forming stronger bonds determines the wrapping direction.

\subsection*{Optimal radius descriptors}

All studied tubes are more stable if the chalcogen element is placed inside. In fact, for some of the studied non-isovalent nanotubes the optimal tube radius is independent of the halogen sitting on the outside of the tube (\textit{e.g.} SbSCl, SbSBr and SbSI). Thus, only considering MX$_2$ and MY$_2$ parent structures provides an inaccurate picture when studying non-isovalent nanotubes. Additionally, the parent structures of non-isovalent tubes will each have a M-element oxidation state different from the 2D Janus sheet (\textit{e.g.} BiI$_2$, BiS$_2$ and BiSI). Two models from literature use the parent structure lattice constants as descriptors for predicting the optimal tube radius of isovalent nanotubes. The first approach is based on the plate theory as described by Timoshenko  \cite{timoshenko1925analysis, xiong2018spontaneous}, while the second approach makes use of the Poisson ratios of the parent sheets (Poisson model) \cite{zhao2015ultra}. Both models require the calculation of the stiffness tensors of the parent structures. Our approach, which we refer to as Inner-bond model, is instead based on solely geometrical parameters, which can be readily obtained from existing 2D databases (\textit{e.g.} \cite{haastrup2018computational}). It uses the lattice constants of the 2D MXY Janus and MX$_2$ parent sheets as well as the 2D Janus t$_{M-X}$ layer thickness, as discussed above. It is initially assumed that the optimal tube radius is determined by the the lattice mismatch between the MXY Janus sheet and the corresponding MX$_{2}$ parent sheet. The derivation of the Inner-bond model and the used formulas for the two reference models are provided in the SI.

\begin{figure}[h]
\centering
\begin{subfigure}{.6\textwidth}
  \centering
  \includegraphics[width=1.\linewidth]{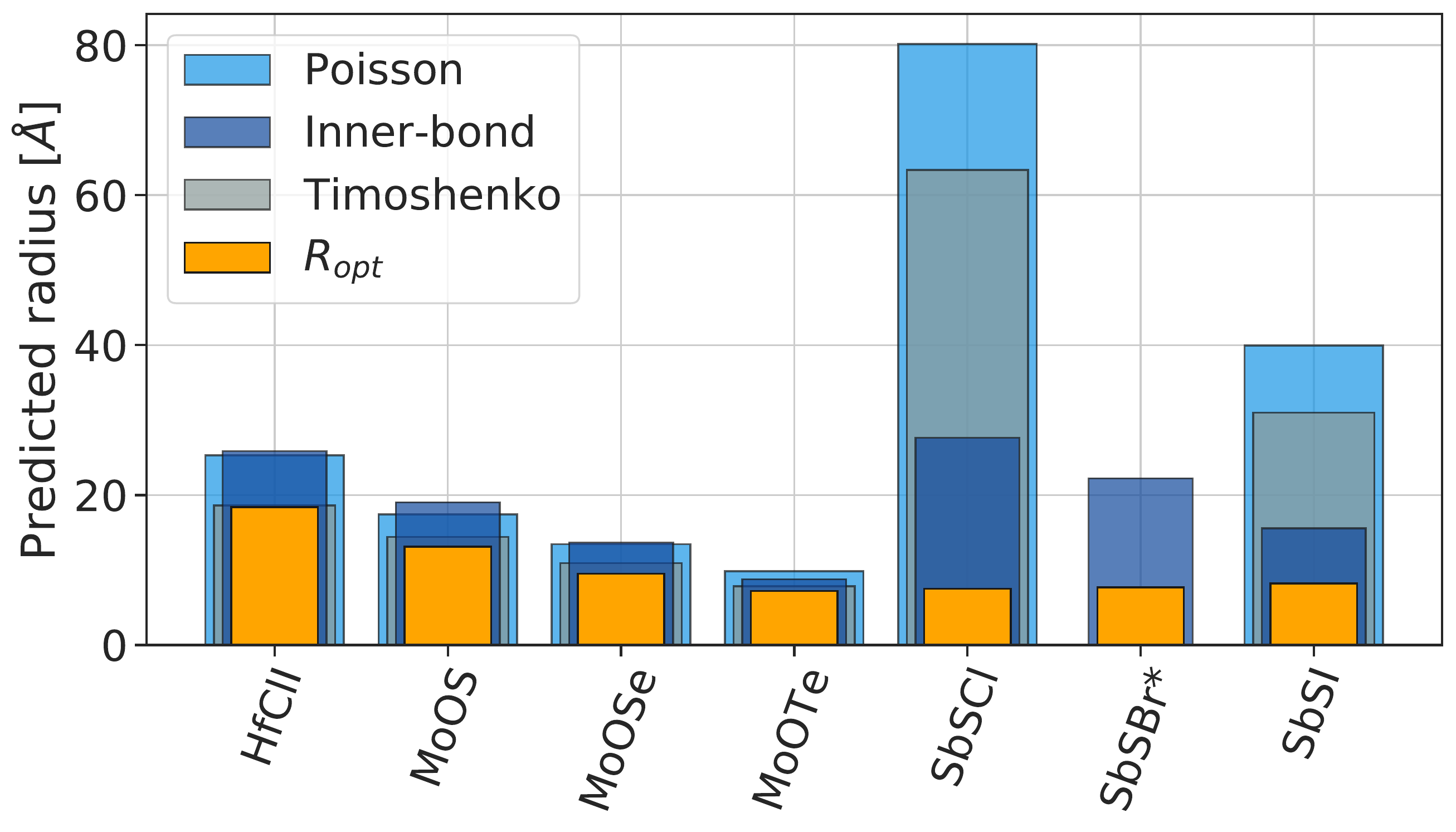}
  \subcaption{}\label{fig:1a}
  \label{fig:compare_simple_models}
\end{subfigure}%
\begin{subfigure}{.4\textwidth}
  \centering
  \includegraphics[width=1.\linewidth]{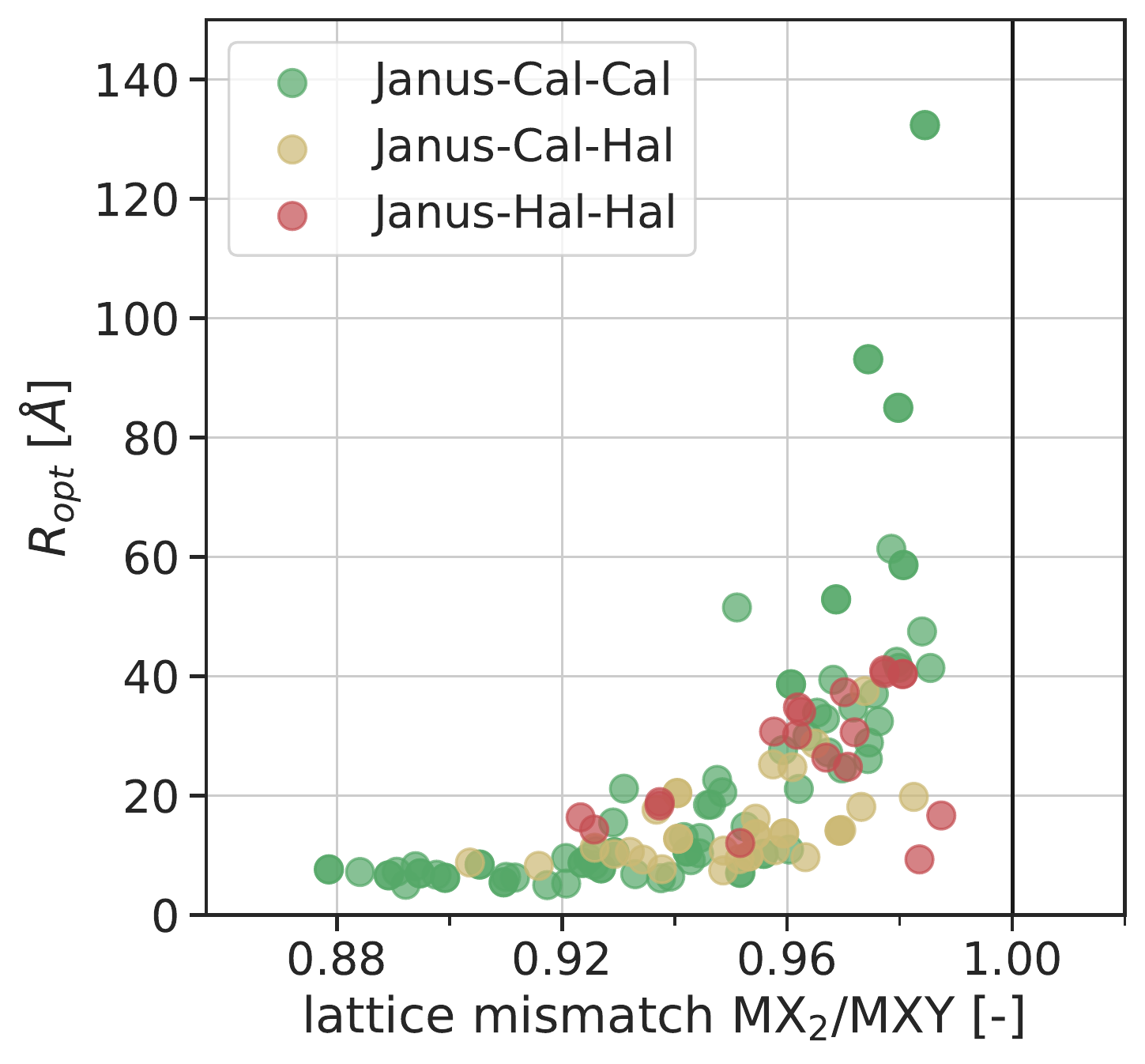}
  \subcaption{}\label{fig:1b}
  \label{fig:tube_radius_predicted}
\end{subfigure}
\caption{ (a) Optimal tube radius using the three models based on the Timoshenko plate theory \cite{timoshenko1925analysis}, considering the Poisson ratios (Poisson) \cite{zhao2015ultra} and using lattice mismatches between Janus MXY and MY$_2$ parent sheets (Inner-bond model). The optimal radius based on DFT calculations R$_{opt}$ is also shown which has been obtained extrapolating the function given in equation \ref{eq:second_order_model}. The asterisk indicates that due to prototype changes when straining the SbBr$_{2}$ sheet, it was not possible to get the elastic tensor required for the Timoshenko and Poisson model. (b) The radius plotted against the lattice mismatch MX$_2$/MXY.}
\label{fig:tube_radius_predicted}
\end{figure}

All three models capture the rolling mechanism for isovalent anions for a selected set of materials (with discrepancies up to 30 \% compared to the calculated DFT radius - Figure \ref{fig:tube_radius_predicted} (a)), while only the Inner-bond model predicts the optimal radii of non-isovalent nanotubes to be small. Nevertheless, it fails to capture that the optimal DFT radius for some tubes, in which a chalcogen sits inside the tube, is almost independent of the chosen halogen element sitting outside of the tube (see orange bars for SbSCl, SbSBr and SbSI in Figure \ref{fig:tube_radius_predicted} (a)). Here, the layer thickness t$_{M-X}$ is so thin, that the steric effects of the halogen atoms at small radii do not decisively impact the stability of the tube. In this picture, the wrapping mechanism is solely governed by the steric effects of the chalcogen elements. Controversely, the 2D Janus sheet lattice constants increases for SbSCl $<$ SbSBr $<$ SbSI due to the size difference of the halogen atoms, leading to a mismatch in predicted values for the Inner-bond model. \\

One feature all the studied materials share is that the lattice mismatch between the parent MX$_2$ and the Janus MXY sheets is always smaller than 1 (Figure \ref{fig:tube_radius_predicted} (b)). This is different from the idea of comparing MX$_2$ and MY$_2$ lattice constants, for which, in the case of mixing chalcogen and halogen X/Y elements, the lattice mismatch exceeds 1 (Figure \ref{fig:properties_correlated} (a)). Although only giving quantitative predictions, the Inner-bond model shows that the bond distances of the elements in the inner MX$_2$ sheet together with the MXY lattice constant are good descriptors for predicting the optimal radius within the studied compound space. In the case of isovalent anions the lattice-mismatch of the parent structures can also be used as a descriptor.

\subsection*{Strain energy descriptors}

\begin{figure}[h]
\centering
  \includegraphics[width=1\linewidth]{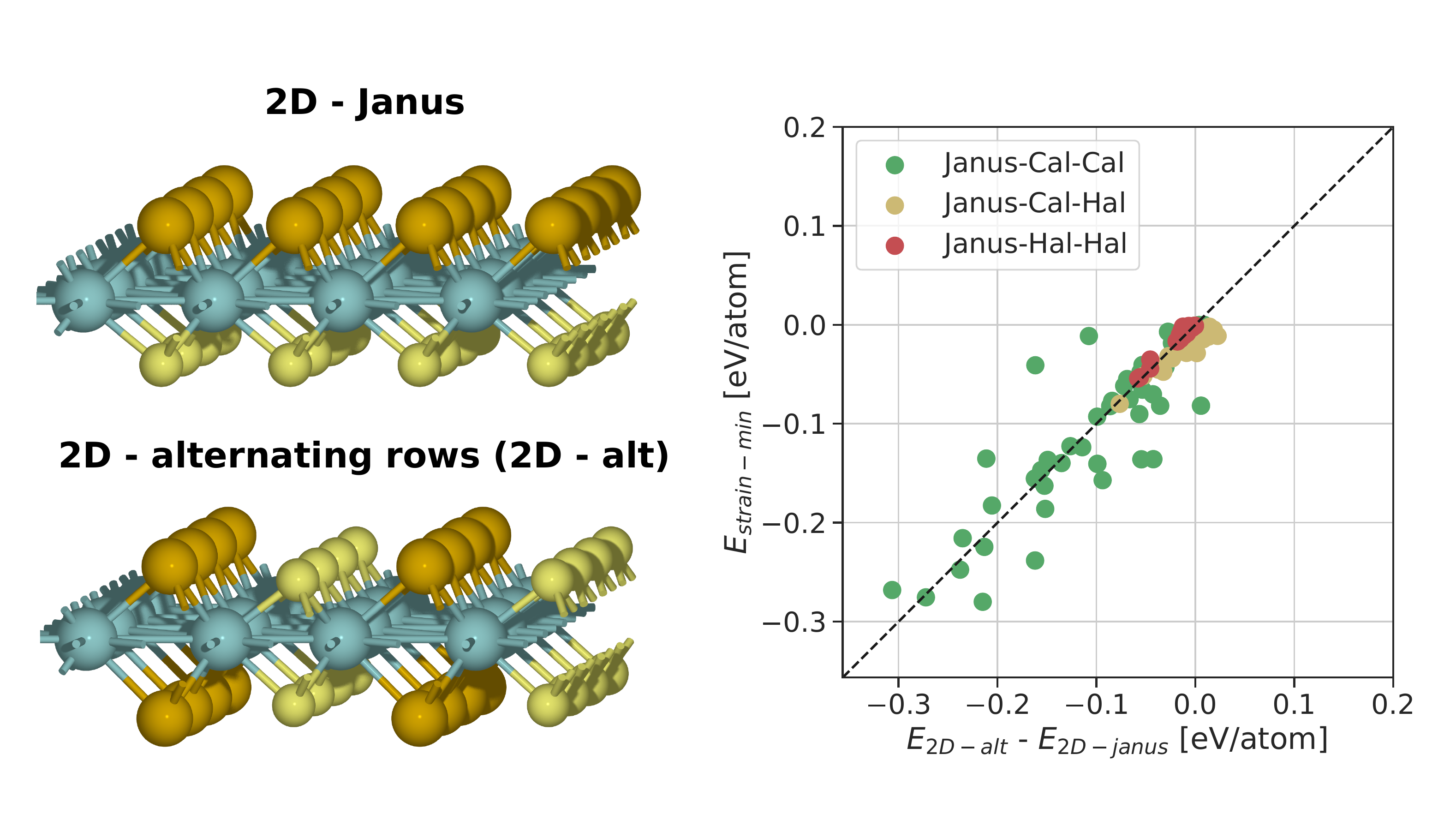}
\caption{Comparing the energy of the tube at its optimal radius versus the energy of a sheet consisting of alternating anion rows (2D-alternating rows prototype structure or 2D-alt). The mean average error for the different combinations are 22.9 meV/atom (Janus-Cal-Cal), 14.4 meV/atom (Janus-Cal-Hal) and 3.1 meV/atom (Janus-Hal-Hal).}
\label{fig:alternating_sheets}
\end{figure}

The Janus sheets show an asymmetry between the two sides of the layer. We can make a sheet where the mirror symmetry along the mid-layer (M-layer) is restored through rearranging the X/Y elements into alternating X/Y rows (as shown in Figure \ref{fig:alternating_sheets}). The graph in Figure \ref{fig:alternating_sheets} shows the energy difference between the alternating and the Janus sheet (E$_{2D-alt}$ - E$_{2D-Janus}$) as a descriptor for the strain energy minumum ($E_{strain-min}$). Few outliers ( $>$ 50 meV/atom error in prediction) correspond to structures with a low convex hull stability, \textit{i.e.} $E_{CH}$ $>$ 0.2 eV/atom.

For instance, this descriptor, which does not require the calculation of the tubes, could be used to indicate if a sheet exfoliated onto a host structure might undergo spontaneous curling. The strain energy would need to be larger than the adsorption energy of the sheet on the host structure,  which is for example not the case for the experimentally exfoliated BiTeI sheet \cite{fulop2018exfoliation}. Whether a large strain energy is needed to form single-wall nanotubes (in favor of multi-wall nanotubes or other forms of rolled structures) requires a deeper experimental investigation, or, at least, to consider growing and other experimental conditions which might impact the formation mechanism.\cite{monet2018structural} 

Different synthesis procedures can be imagined leading to the synthesis of the Janus nanotubes suggested here. Besides the exfoliation of a monolayer onto a host structure and its subsequent spontaneous wrapping, which might need lithography techniques to cut the monolayer into the required dimensions, the intercalation of ions into Janus multilayered materials, combined with ball-milling, \cite{jung2016intercalation} can be a viable experimental synthesis path allowing to produce larger quantities. A common issue with these methods would be preventing unwanted reactions occurring at unsaturated bonds. A possible alternative synthesis pathway could be to use Atomic Layer Deposition combined with lithography to cut the 2D layer into a desired shape before it spontaneously wraps. Regardless the synthesis procedure, the intrinsic driving force of the 2D Janus sheets to form 1D tubes (or alternatively other curled shapes) should be large enough to be observed.

A promising starting material to experimentally verify our predictions is the layered BiTeI 3D bulk crystal structure for which experimental synthesis routes have been established.\cite{shevelkov1995crystal} Other interesting candidates include tubes based on Vanadium or Titanium, which, by displaying a variety of oxidation states, are suitable for electrochemical applications.\cite{liu2007vanadium, kianfar2019recent} Curled multiwall VO$_x$ structures with much larger inner radii than the tubes predicted here have already been reported.\cite{kianfar2019recent} Other non-isovalent anion based tubes, such as BiSI, might be feasible. However, the weaker ionic halogen bonds might drive reconstructions, in contrast to less reactive covalent bonds.

\section*{Discussion}
In this work, we investigated the formation mechanism of nanotubes by studying their stability and optimal nanotube radius. 135 different Janus nanotubes have been calculated using a structure prototype approach (T-/H-phase). Each three-layered material consisted of one of 15 different cation mid-layer elements in combination with inner and outer atomic layers being occupied by either chalcogens, halogens or their mixture as anions. For isovalent anions, the wrapping mechanism could be explained by the lattice-mismatch between the two inner and outer atomic layers, while for non-isovalent anions steric effects caused through short pnictogen-chalcogen inside and longer pnictogen-halogen bonds outside of the material drive the stability. These effects are beneficial to the formation of some of the smallest identified nanotubes showing optimal radii below 10 \si{\angstrom}. We noted that in general a large minimum strain energy is not needed to find tubes with an optimal radius smaller than 35 \si{\angstrom}. Additionally, the minimum strain energy was reasonably well estimated using the energy difference between a 2D Janus and alternating sheet as a descriptor. We employed Bayesian statistics to assess the quality of our fitting in order to identify uncertainties in our predictions of the optimal radius due, for example, to a misfit between the obtained data and the underlying strain energy curve function. 

Nanotubes based on BiTeI and related compositions appear to be a particular interesting starting point for experimental verification due to their synthesizability in a 3D structure and exfoliability into 2D layers, as well as their predicted stability in 1D form. Following the interest in Vanadium- and Titanium-based nanotubes for electrochemical applications, we suggest that the metastable combinations of these metals paired with either OTe, OSe, or OS can be additional interesting candidates.

The findings reported here, shed light on the mechanism behind the curling of 2D Janus sheets and define a new path for the synthesis of nanotubes with small radii, for which the lattice mismatch and the bonding character of the anions play a fundamental role.

\section*{Methods}
The first step to create our library of nanotubes is to relax the 2D Janus sheets, taken from the Computational 2D Materials Database (c2db).\cite{haastrup2018computational} If a 2D Janus sheet is not present in the database, the 2D Janus sheet is created by averaging the lattice constants from its constituent MX$_{2}$ and MY$_{2}$ parent sheets. Subsequently, tubes are generated by repeating and wrapping the 2D sheets both along the armchair and zigzag wrapping directions, thereby obtaining tubes with various radii (similar to what is shown in Figure \ref{fig:wrap_tube_figure}). In details, the initial number of unit cell repetitions is n = (6, 8, 10, 12, 14) for the armchair  and n = (10, 13, 16, 19, 22) for the zigzag wrapping direction, which correspond to tubes with a radius smaller than 20 \si{\angstrom}. We apply a set of filters to decide whether or not the relaxed structure is accepted for further investigation. These filters include assuring that a tube retained its circular shape and that no unwanted changes into different prototypes occurred during the relaxation. The filters discard $\sim$ 40\% of the data generated. A detailed discussion on how the data is filtered prior to visualization can be found in the SI. For consistent and reproducible calculations, we implement a workflow combining the Atomic Simulation Environment (ASE) \cite{larsen2017atomic} and the workflow scheduling system MyQueue \cite{mortensen2020myqueue}. Inspired by the CUSTODIAN package \cite{ong2013python}, we establish an "ASE error handler" to handle common DFT errors thus limiting the need for user intervention. A similar approach has been recently implemented to autonomously discover battery electrodes.\cite{bolle2019autonomous}

All calculations are carried out with the Vienna ab initio Simulation Package (VASP) using a plane-wave basis set with an energy cutoff of 550 eV. \cite{kresse1996efficient,kresse1999ultrasoft,blochl1994projector} In order to approximate the exchange-correlation effects the Perdew-Burke-Ernzerhof (PBE) form generalized gradient approximation (GGA) is used. \cite{perdew1996generalized} A k-point density $>$ 4.7 $/\si{\angstrom}^{-1}$ is used to sample the Brillouin zone. All forces are converged to less than 0.01 eV/\si{\angstrom}. The structures are relaxed in a non-magnetic configuration, \textit{i.e.} without applying initial magnetic moments on the elements. A minimum vacuum in between repeating images of 16 \si{\angstrom} is ensured. Dipole corrections are applied along the non-periodic direction for materials with an out-of-plane dipole moment. To assess the stability versus 3D phases, we use a convex hull analysis, where the reference structures are the ones defining the convex hull in the Materials Project database. \cite{jain2013commentary} These structures are then relaxed with the matching input parameters used in this work, \cite{ong2013python} while the reference energy of oxygen is obtained by calculating the difference in energy between water and hydrogen in the gas phase, including the zero point energy (ZPE) corrections. \cite{rossmeisl2005electrolysis} 

\subsection*{Code availability}
The workflow is continuously developed and may be accessed at https://gitlab.com/ivca/Nanotubes.

\section*{Data availability}
The structures presented in this work are available on DTU DATA with the identifier "doi.org/10.11583/DTU.13153355".



\section*{Acknowledgements}
The authors thank Prof. Poul Norby for the valuable discussion on the possible synthesis pathways of the investigated tubes. The authors acknowledge support from the Department of Energy Conversion and Storage, Technical University of Denmark, through the Special Competence Initiative Autonomous Materials Discovery (AiMade - http://aimade.org/). KST acknowledges support from the European Research Council (ERC) under the European Union’s Horizon 2020 research and innovation programme (grant agreement No 773122, LIMA).

\section*{Supplementary Information}

Filtering of the database, formulas used to predict the optimal radius, graphs include a comparison of BiSI nanotubes to experimental reference tubes, the optimal radius versus the convex hull stability and a comparison of tube energies versus layer thicknesses

\section*{Author contributions}
F.T.B. wrote all scripts for the analysis of the data and the draft of the manuscript.
F.T.B and A.E.G.M. implemented the workflow and performed the calculations.
T.V. and I.E.C. designed the project.
All authors contributed to discussing the results as well as writing and revising of the manuscript.

\section*{Competing interests}
The authors declare no competing interests.


\begin{thebibliography}{10}
\expandafter\ifx\csname url\endcsname\relax
  \def\url#1{\texttt{#1}}\fi
\expandafter\ifx\csname urlprefix\endcsname\relax\def\urlprefix{URL }\fi
\providecommand{\bibinfo}[2]{#2}
\providecommand{\eprint}[2][]{\url{#2}}

\bibitem{serra2019overview}
\bibinfo{author}{Serra, M.}, \bibinfo{author}{Arenal, R.} \&
  \bibinfo{author}{Tenne, R.}
\newblock \bibinfo{title}{An overview of the recent advances in inorganic
  nanotubes}.
\newblock \emph{\bibinfo{journal}{Nanoscale}} \textbf{\bibinfo{volume}{11}},
  \bibinfo{pages}{8073--8090} (\bibinfo{year}{2019}).

\bibitem{iijima1991helical}
\bibinfo{author}{Iijima, S.}
\newblock \bibinfo{title}{Helical microtubules of graphitic carbon}.
\newblock \emph{\bibinfo{journal}{nature}} \textbf{\bibinfo{volume}{354}},
  \bibinfo{pages}{56--58} (\bibinfo{year}{1991}).

\bibitem{Tenne1992}
\bibinfo{author}{Tenne, R.}, \bibinfo{author}{Margulis, L.},
  \bibinfo{author}{Genut, M.} \& \bibinfo{author}{Hodes, G.}
\newblock \bibinfo{title}{Polyhedral and cylindrical structures of tungsten
  disulphide}.
\newblock \emph{\bibinfo{journal}{Nature}} \textbf{\bibinfo{volume}{360}},
  \bibinfo{pages}{444--446} (\bibinfo{year}{1992}).
\newblock \urlprefix\url{https://doi.org/10.1038/360444a0}.

\bibitem{rao2003inorganic}
\bibinfo{author}{Rao, C. N.~R.} \& \bibinfo{author}{Nath, M.}
\newblock \bibinfo{title}{Inorganic nanotubes}.
\newblock In \emph{\bibinfo{booktitle}{Advances In Chemistry: A Selection of
  CNR Rao's Publications (1994--2003)}}, \bibinfo{pages}{310--333}
  (\bibinfo{publisher}{World Scientific}, \bibinfo{year}{2003}).

\bibitem{seifert2002stability}
\bibinfo{author}{Seifert, G.}, \bibinfo{author}{K{\"o}hler, T.} \&
  \bibinfo{author}{Tenne, R.}
\newblock \bibinfo{title}{Stability of metal chalcogenide nanotubes}.
\newblock \emph{\bibinfo{journal}{The Journal of Physical Chemistry B}}
  \textbf{\bibinfo{volume}{106}}, \bibinfo{pages}{2497--2501}
  (\bibinfo{year}{2002}).

\bibitem{pauling1930structure}
\bibinfo{author}{Pauling, L.}
\newblock \bibinfo{title}{The structure of the chlorites}.
\newblock \emph{\bibinfo{journal}{Proceedings of the National Academy of
  Sciences of the United States of America}} \textbf{\bibinfo{volume}{16}},
  \bibinfo{pages}{578} (\bibinfo{year}{1930}).

\bibitem{yoshinaga1962imogolite}
\bibinfo{author}{Yoshinaga, N.} \& \bibinfo{author}{Aomine, S.}
\newblock \bibinfo{title}{Imogolite in some ando soils}.
\newblock \emph{\bibinfo{journal}{Soil Science and Plant Nutrition}}
  \textbf{\bibinfo{volume}{8}}, \bibinfo{pages}{22--29} (\bibinfo{year}{1962}).

\bibitem{cradwick1972imogolite}
\bibinfo{author}{Cradwick, P.} \emph{et~al.}
\newblock \bibinfo{title}{Imogolite, a hydrated aluminium silicate of tubular
  structure}.
\newblock \emph{\bibinfo{journal}{Nature Physical Science}}
  \textbf{\bibinfo{volume}{240}}, \bibinfo{pages}{187--189}
  (\bibinfo{year}{1972}).

\bibitem{farmer1977synthesis}
\bibinfo{author}{Farmer, V.}, \bibinfo{author}{Fraser, A.} \&
  \bibinfo{author}{Tait, J.}
\newblock \bibinfo{title}{Synthesis of imogolite: a tubular aluminium silicate
  polymer}  (\bibinfo{year}{1977}).

\bibitem{bates1950morphology}
\bibinfo{author}{Bates, T.~F.}, \bibinfo{author}{Hildebrand, F.~A.} \&
  \bibinfo{author}{Swineford, A.}
\newblock \bibinfo{title}{Morphology and structure of endellite and
  halloysite}.
\newblock \emph{\bibinfo{journal}{American Mineralogist: Journal of Earth and
  Planetary Materials}} \textbf{\bibinfo{volume}{35}},
  \bibinfo{pages}{463--484} (\bibinfo{year}{1950}).

\bibitem{whittaker1956structure}
\bibinfo{author}{Whittaker, E.}
\newblock \bibinfo{title}{The structure of chrysotile. ii. clino-chrysotile}.
\newblock \emph{\bibinfo{journal}{Acta Crystallographica}}
  \textbf{\bibinfo{volume}{9}}, \bibinfo{pages}{855--862}
  (\bibinfo{year}{1956}).

\bibitem{panchakarla2014nanotubes}
\bibinfo{author}{Panchakarla, L.~S.} \emph{et~al.}
\newblock \bibinfo{title}{Nanotubes from misfit layered compounds: A new family
  of materials with low dimensionality}.
\newblock \emph{\bibinfo{journal}{The journal of physical chemistry letters}}
  \textbf{\bibinfo{volume}{5}}, \bibinfo{pages}{3724--3736}
  (\bibinfo{year}{2014}).


\bibitem{august2020}
\bibinfo{author}{Mikkelsen, A. E. G.} \emph{et~al.}
\newblock \bibinfo{title}{Band Structure of MoSTe Janus Nanotubes}.
\newblock \emph{\bibinfo{journal}{Under review}}
  (\bibinfo{year}{2020}).
 
\bibitem{lu2017janus}
\bibinfo{author}{Lu, A.-Y.} \emph{et~al.}
\newblock \bibinfo{title}{Janus monolayers of transition metal
  dichalcogenides}.
\newblock \emph{\bibinfo{journal}{Nature nanotechnology}}
  \textbf{\bibinfo{volume}{12}}, \bibinfo{pages}{744--749}
  (\bibinfo{year}{2017}).

\bibitem{fulop2018exfoliation}
\bibinfo{author}{F{\"u}l{\"o}p, B.} \emph{et~al.}
\newblock \bibinfo{title}{Exfoliation of single layer bitei flakes}.
\newblock \emph{\bibinfo{journal}{2D Materials}} \textbf{\bibinfo{volume}{5}},
  \bibinfo{pages}{031013} (\bibinfo{year}{2018}).

\bibitem{zhao2015ultra}
\bibinfo{author}{Zhao, W.}, \bibinfo{author}{Li, Y.}, \bibinfo{author}{Duan,
  W.} \& \bibinfo{author}{Ding, F.}
\newblock \bibinfo{title}{Ultra-stable small diameter hybrid transition metal
  dichalcogenide nanotubes x--m--y (x, y= s, se, te; m= mo, w, nb, ta): a
  computational study}.
\newblock \emph{\bibinfo{journal}{Nanoscale}} \textbf{\bibinfo{volume}{7}},
  \bibinfo{pages}{13586--13590} (\bibinfo{year}{2015}).

\bibitem{oshima2020geometrical}
\bibinfo{author}{Oshima, S.}, \bibinfo{author}{Toyoda, M.} \&
  \bibinfo{author}{Saito, S.}
\newblock \bibinfo{title}{Geometrical and electronic properties of unstrained
  and strained transition metal dichalcogenide nanotubes}.
\newblock \emph{\bibinfo{journal}{Physical Review Materials}}
  \textbf{\bibinfo{volume}{4}}, \bibinfo{pages}{026004} (\bibinfo{year}{2020}).

\bibitem{riis2019classifying}
\bibinfo{author}{Riis-Jensen, A.~C.}, \bibinfo{author}{Deilmann, T.},
  \bibinfo{author}{Olsen, T.} \& \bibinfo{author}{Thygesen, K.~S.}
\newblock \bibinfo{title}{Classifying the electronic and optical properties of
  janus monolayers}.
\newblock \emph{\bibinfo{journal}{ACS nano}} \textbf{\bibinfo{volume}{13}},
  \bibinfo{pages}{13354--13364} (\bibinfo{year}{2019}).

\bibitem{tersoff1992energies}
\bibinfo{author}{Tersoff, J.}
\newblock \bibinfo{title}{Energies of fullerenes}.
\newblock \emph{\bibinfo{journal}{Physical Review B}}
  \textbf{\bibinfo{volume}{46}}, \bibinfo{pages}{15546} (\bibinfo{year}{1992}).

\bibitem{tibbetts1984carbon}
\bibinfo{author}{Tibbetts, G.~G.}
\newblock \bibinfo{title}{Why are carbon filaments tubular?}
\newblock \emph{\bibinfo{journal}{Journal of crystal growth}}
  \textbf{\bibinfo{volume}{66}}, \bibinfo{pages}{632--638}
  (\bibinfo{year}{1984}).

\bibitem{robertson1992energetics}
\bibinfo{author}{Robertson, D.}, \bibinfo{author}{Brenner, D.} \&
  \bibinfo{author}{Mintmire, J.}
\newblock \bibinfo{title}{Energetics of nanoscale graphitic tubules}.
\newblock \emph{\bibinfo{journal}{Physical Review B}}
  \textbf{\bibinfo{volume}{45}}, \bibinfo{pages}{12592} (\bibinfo{year}{1992}).

\bibitem{guimaraes2007imogolite}
\bibinfo{author}{Guimar{\~a}es, L.} \emph{et~al.}
\newblock \bibinfo{title}{Imogolite nanotubes: stability, electronic, and
  mechanical properties}.
\newblock \emph{\bibinfo{journal}{Acs Nano}} \textbf{\bibinfo{volume}{1}},
  \bibinfo{pages}{362--368} (\bibinfo{year}{2007}).

\bibitem{d2009single}
\bibinfo{author}{D’Arco, P.}, \bibinfo{author}{Noel, Y.},
  \bibinfo{author}{Demichelis, R.} \& \bibinfo{author}{Dovesi, R.}
\newblock \bibinfo{title}{Single-layered chrysotile nanotubes: A quantum
  mechanical ab initio simulation}.
\newblock \emph{\bibinfo{journal}{The Journal of chemical physics}}
  \textbf{\bibinfo{volume}{131}}, \bibinfo{pages}{204701}
  (\bibinfo{year}{2009}).

\bibitem{guimaraes2010structural}
\bibinfo{author}{Guimaraes, L.}, \bibinfo{author}{Enyashin, A.~N.},
  \bibinfo{author}{Seifert, G.} \& \bibinfo{author}{Duarte, H.~A.}
\newblock \bibinfo{title}{Structural, electronic, and mechanical properties of
  single-walled halloysite nanotube models}.
\newblock \emph{\bibinfo{journal}{The Journal of Physical Chemistry C}}
  \textbf{\bibinfo{volume}{114}}, \bibinfo{pages}{11358--11363}
  (\bibinfo{year}{2010}).

\bibitem{zhang2003formation}
\bibinfo{author}{Zhang, S.} \emph{et~al.}
\newblock \bibinfo{title}{Formation mechanism of h 2 t i 3 o 7 nanotubes}.
\newblock \emph{\bibinfo{journal}{Physical Review Letters}}
  \textbf{\bibinfo{volume}{91}}, \bibinfo{pages}{256103}
  (\bibinfo{year}{2003}).

\bibitem{demichelis2016serpentine}
\bibinfo{author}{Demichelis, R.}, \bibinfo{author}{De~La~Pierre, M.},
  \bibinfo{author}{Mookherjee, M.}, \bibinfo{author}{Zicovich-Wilson, C.~M.} \&
  \bibinfo{author}{Orlando, R.}
\newblock \bibinfo{title}{Serpentine polymorphism: a quantitative insight from
  first-principles calculations}.
\newblock \emph{\bibinfo{journal}{CrystEngComm}} \textbf{\bibinfo{volume}{18}},
  \bibinfo{pages}{4412--4419} (\bibinfo{year}{2016}).

\bibitem{ong2013python}
\bibinfo{author}{Ong, S.~P.} \emph{et~al.}
\newblock \bibinfo{title}{Python materials genomics (pymatgen): A robust,
  open-source python library for materials analysis}.
\newblock \emph{\bibinfo{journal}{Computational Materials Science}}
  \textbf{\bibinfo{volume}{68}}, \bibinfo{pages}{314--319}
  (\bibinfo{year}{2013}).

\bibitem{castelli2012computational}
\bibinfo{author}{Castelli, I.~E.} \emph{et~al.}
\newblock \bibinfo{title}{Computational screening of perovskite metal oxides
  for optimal solar light capture}.
\newblock \emph{\bibinfo{journal}{Energy \& Environmental Science}}
  \textbf{\bibinfo{volume}{5}}, \bibinfo{pages}{5814--5819}
  (\bibinfo{year}{2012}).

\bibitem{sun2016thermodynamic}
\bibinfo{author}{Sun, W.} \emph{et~al.}
\newblock \bibinfo{title}{The thermodynamic scale of inorganic crystalline
  metastability}.
\newblock \emph{\bibinfo{journal}{Science advances}}
  \textbf{\bibinfo{volume}{2}}, \bibinfo{pages}{e1600225}
  (\bibinfo{year}{2016}).

\bibitem{esposito2020metastability}
\bibinfo{author}{Esposito, V.} \& \bibinfo{author}{Castelli, I.~E.}
\newblock \bibinfo{title}{Metastability at defective metal oxide interfaces and
  nanoconfined structures}.
\newblock \emph{\bibinfo{journal}{Advanced Materials Interfaces}}
  \bibinfo{pages}{1902090} (\bibinfo{year}{2020}).

\bibitem{lee2011origin}
\bibinfo{author}{Lee, S.~U.}, \bibinfo{author}{Choi, Y.~C.},
  \bibinfo{author}{Youm, S.~G.} \& \bibinfo{author}{Sohn, D.}
\newblock \bibinfo{title}{Origin of the strain energy minimum in imogolite
  nanotubes}.
\newblock \emph{\bibinfo{journal}{The Journal of Physical Chemistry C}}
  \textbf{\bibinfo{volume}{115}}, \bibinfo{pages}{5226--5231}
  (\bibinfo{year}{2011}).

\bibitem{shevelkov1995crystal}
\bibinfo{author}{Shevelkov, A.}, \bibinfo{author}{Dikarev, E.},
  \bibinfo{author}{Shpanchenko, R.} \& \bibinfo{author}{Popovkin, B.}
\newblock \bibinfo{title}{Crystal structures of bismuth tellurohalides bitex
  (x= cl, br, i) from x-ray powder diffraction data}.
\newblock \emph{\bibinfo{journal}{Journal of Solid State Chemistry}}
  \textbf{\bibinfo{volume}{114}}, \bibinfo{pages}{379--384}
  (\bibinfo{year}{1995}).

\bibitem{timoshenko1925analysis}
\bibinfo{author}{Timoshenko, S.}
\newblock \bibinfo{title}{Analysis of bi-metal thermostats}.
\newblock \emph{\bibinfo{journal}{Josa}} \textbf{\bibinfo{volume}{11}},
  \bibinfo{pages}{233--255} (\bibinfo{year}{1925}).

\bibitem{xiong2018spontaneous}
\bibinfo{author}{Xiong, Q.-l.}, \bibinfo{author}{Zhou, J.},
  \bibinfo{author}{Zhang, J.}, \bibinfo{author}{Kitamura, T.} \&
  \bibinfo{author}{Li, Z.-h.}
\newblock \bibinfo{title}{Spontaneous curling of freestanding janus monolayer
  transition-metal dichalcogenides}.
\newblock \emph{\bibinfo{journal}{Physical Chemistry Chemical Physics}}
  \textbf{\bibinfo{volume}{20}}, \bibinfo{pages}{20988--20995}
  (\bibinfo{year}{2018}).

\bibitem{haastrup2018computational}
\bibinfo{author}{Haastrup, S.} \emph{et~al.}
\newblock \bibinfo{title}{The computational 2d materials database:
  high-throughput modeling and discovery of atomically thin crystals}.
\newblock \emph{\bibinfo{journal}{2D Materials}} \textbf{\bibinfo{volume}{5}},
  \bibinfo{pages}{042002} (\bibinfo{year}{2018}).

\bibitem{monet2018structural}
\bibinfo{author}{Monet, G.} \emph{et~al.}
\newblock \bibinfo{title}{Structural resolution of inorganic nanotubes with
  complex stoichiometry}.
\newblock \emph{\bibinfo{journal}{Nature communications}}
  \textbf{\bibinfo{volume}{9}}, \bibinfo{pages}{1--9} (\bibinfo{year}{2018}).

\bibitem{jung2016intercalation}
\bibinfo{author}{Jung, Y.}, \bibinfo{author}{Zhou, Y.} \& \bibinfo{author}{Cha,
  J.~J.}
\newblock \bibinfo{title}{Intercalation in two-dimensional transition metal
  chalcogenides}.
\newblock \emph{\bibinfo{journal}{Inorganic Chemistry Frontiers}}
  \textbf{\bibinfo{volume}{3}}, \bibinfo{pages}{452--463}
  (\bibinfo{year}{2016}).

\bibitem{liu2007vanadium}
\bibinfo{author}{Liu, A.}, \bibinfo{author}{Ichihara, M.},
  \bibinfo{author}{Honma, I.} \& \bibinfo{author}{Zhou, H.}
\newblock \bibinfo{title}{Vanadium-oxide nanotubes: Synthesis and
  template-related electrochemical properties}.
\newblock \emph{\bibinfo{journal}{Electrochemistry communications}}
  \textbf{\bibinfo{volume}{9}}, \bibinfo{pages}{1766--1771}
  (\bibinfo{year}{2007}).

\bibitem{kianfar2019recent}
\bibinfo{author}{Kianfar, E.}
\newblock \bibinfo{title}{Recent advances in synthesis, properties, and
  applications of vanadium oxide nanotube}.
\newblock \emph{\bibinfo{journal}{Microchemical Journal}}
  \textbf{\bibinfo{volume}{145}}, \bibinfo{pages}{966--978}
  (\bibinfo{year}{2019}).

\bibitem{larsen2017atomic}
\bibinfo{author}{Larsen, A.~H.} \emph{et~al.}
\newblock \bibinfo{title}{The atomic simulation environment—a python library
  for working with atoms}.
\newblock \emph{\bibinfo{journal}{Journal of Physics: Condensed Matter}}
  \textbf{\bibinfo{volume}{29}}, \bibinfo{pages}{273002}
  (\bibinfo{year}{2017}).

\bibitem{mortensen2020myqueue}
\bibinfo{author}{Mortensen, J.}, \bibinfo{author}{Gjerding, M.} \&
  \bibinfo{author}{Thygesen, K.}
\newblock \bibinfo{title}{Myqueue: Task and workflow scheduling system}.
\newblock \emph{\bibinfo{journal}{Journal of Open Source Software}}
  \textbf{\bibinfo{volume}{5}}, \bibinfo{pages}{1844} (\bibinfo{year}{2020}).

\bibitem{bolle2019autonomous}
\bibinfo{author}{Bolle, F.~T.} \emph{et~al.}
\newblock \bibinfo{title}{Autonomous discovery of materials for intercalation
  electrodes}.
\newblock \emph{\bibinfo{journal}{Batteries \& Supercaps}}
  (\bibinfo{year}{2019}).

\bibitem{kresse1996efficient}
\bibinfo{author}{Kresse, G.} \& \bibinfo{author}{Furthm{\"u}ller, J.}
\newblock \bibinfo{title}{Efficient iterative schemes for ab initio
  total-energy calculations using a plane-wave basis set}.
\newblock \emph{\bibinfo{journal}{Physical review B}}
  \textbf{\bibinfo{volume}{54}}, \bibinfo{pages}{11169} (\bibinfo{year}{1996}).

\bibitem{kresse1999ultrasoft}
\bibinfo{author}{Kresse, G.} \& \bibinfo{author}{Joubert, D.}
\newblock \bibinfo{title}{From ultrasoft pseudopotentials to the projector
  augmented-wave method}.
\newblock \emph{\bibinfo{journal}{Physical Review B}}
  \textbf{\bibinfo{volume}{59}}, \bibinfo{pages}{1758} (\bibinfo{year}{1999}).

\bibitem{blochl1994projector}
\bibinfo{author}{Bl{\"o}chl, P.~E.}
\newblock \bibinfo{title}{Projector augmented-wave method}.
\newblock \emph{\bibinfo{journal}{Physical review B}}
  \textbf{\bibinfo{volume}{50}}, \bibinfo{pages}{17953} (\bibinfo{year}{1994}).

\bibitem{perdew1996generalized}
\bibinfo{author}{Perdew, J.~P.}, \bibinfo{author}{Burke, K.} \&
  \bibinfo{author}{Ernzerhof, M.}
\newblock \bibinfo{title}{Generalized gradient approximation made simple}.
\newblock \emph{\bibinfo{journal}{Physical review letters}}
  \textbf{\bibinfo{volume}{77}}, \bibinfo{pages}{3865} (\bibinfo{year}{1996}).

\bibitem{jain2013commentary}
\bibinfo{author}{Jain, A.} \emph{et~al.}
\newblock \bibinfo{title}{Commentary: The materials project: A materials genome
  approach to accelerating materials innovation}.
\newblock \emph{\bibinfo{journal}{Apl Materials}} \textbf{\bibinfo{volume}{1}},
  \bibinfo{pages}{011002} (\bibinfo{year}{2013}).

\bibitem{rossmeisl2005electrolysis}
\bibinfo{author}{Rossmeisl, J.}, \bibinfo{author}{Logadottir, A.} \&
  \bibinfo{author}{N{\o}rskov, J.~K.}
\newblock \bibinfo{title}{Electrolysis of water on (oxidized) metal surfaces}.
\newblock \emph{\bibinfo{journal}{Chemical physics}}
  \textbf{\bibinfo{volume}{319}}, \bibinfo{pages}{178--184}
  (\bibinfo{year}{2005}).

\end{thebibliography}
\end{document}